\documentclass[3pt,onecolumn]{article}
\usepackage{graphicx}

\newcommand{\ben}{\begin{enumerate}}
\newcommand{\een}{\end{enumerate}}

\newcommand{\be}{\begin{equation}}
\newcommand{\ee}{\end{equation}}
\newcommand{\bea}{\begin{eqnarray}}
\newcommand{\eea}{\end{eqnarray}}

\begin{document}
\begin{flushright}
\end{flushright}
\vspace{0.1cm}
\thispagestyle{empty}

\begin{center}
{\Large\bf 
Hidden Beauty in Twisted Viking Neck Rings
}\\[13mm]
{\rm Kasper Olsen\footnote{The Department of Physics, Technical University of Denmark, Building 307, 2800 Kongens Lyngby, Denmark.  E-mail: kasper.olsen@fysik.dtu.dk} and Jakob Bohr\footnote{Fuel Cell and Solid State Chemistry Division, Ris{\o} DTU, Technical University of Denmark, 4000 Roskilde, Denmark. E-mail: jabo@risoe.dtu.dk}}\\[18mm]

\end{center}

\noindent Hoards found in Ireland, Scotland, Orkney Islands, and Scandinavia 
have revealed the vikings ability to fabricate beautiful arm and neck rings of twisted silver and gold rods \cite{anderson1880,sheehan1991,graham1995,jorgensen1998}. 
Little is known about the fabrication methods of these rings.
Characteristic is the uniform appearance of the twisted pattern along the length of the ring, as well as 
from one ring to another, even when found at different geographical locations. 
How can the twisted pattern displayed be so perfectly repetitive -- what craftsman skills did it demand of the vikings?

We suggest that the vikings utilized a self-forming motif, and demonstrate that the rings are maximally rotated structures. The utilization of the self-forming motif then explains the near perfect appearance of the twisted wires. Under these circumstances, the twisted rings approach a universal geometry which only depends on the number of wires in the material. 

There is a variety of styles in which the viking arm and neck rings were fabricated. Three common patterns are: (i) Thin intertwined metal strands fabricated from (usually two or three) wires of constant thickness. A number of such metal strands are then laid next to each other to form the major motif of the ring (Fig. 1A: R1, R4, R5). (ii) The motif is formed from a single twist of wires. In this case the wires are relatively thick and typically constructed with slowly varying thickness. 
(iii) The motif of thick twisted wires is embellished by a thin wire or a thin twisted wire (Fig. 1A: R2, R3).

The existence of a unique self-forming motif can be understood from the study of twisted wires. 
Imagine one wants to make a piece of jewelry from two wires.
The length, $L$, of the twisted structure depends on how much one twists the wires into helical lines, i.e. the length is a function of the number, $n$, of rotations in the material. For $n=0$ the length $L(0)$ is simply that of the individual straight wires. When one twist these two wires, the resulting double strand gets shorter and shorter (Fig. 1B). At a certain point, no more rotations can be added without deforming the material or introducing intrusions -- this is where the curve changes from blue to red. The maximally rotated geometry is universal and therefore independent of the skills of the craftsman. 

The geometry of the maximally twisted wires are characterized by their pitch angle, $v_\bot$, and their number of strands, $N$. The pitch angle is the angle that the helical lines make with the horizontal axis; 
an untwisted straight wire has a pitch angle of $90^\circ$. Table 1 lists the pitch angles of maximally twisted wires and the reduced aspect ratio, $A_r=H/(N D)$, where $H$ is the pitch height and $D$ is the outer diameter. $H$ is measured along the twisted structure and is the distance over which a single strand comes back to its original position. 

Are the arm and neck rings found at various sites maximally twisted? To answer this question, we determined the pitch angles of a number of viking rings. However, to directly measure the pitch angle with reasonable accuracy is difficult. Instead we measured $H$ and $D$ of the twisted helix. From this we determined the reduced aspect ratio, $A_r$, which is directly related to the pitch angle (Fig. 1C). The calculation of $A_r(v_\bot)$ follows the method outlined in refs. \cite{przybyl2001,neukirch2002,olsen2009}. 

In Figure 1C is marked the measurements of ten arm and neck rings (Table 2). Six data points ($\circ$) 
lie close to the maximally rotated structure on the larger side (blue curve). Only ideally moldable tubes would be able to reach the maximally rotated states. Any real material has material properties that limit the number of rotations before the mathematical limit is reached. The clustering of the data points about a pitch angle that is $\sim 5^\circ$ higher shows that this is what in practice was obtained, leading to a nearly perfectly repeating pattern. For comparison, four rings ($\bullet$) which are not maximally rotated are included.

The above analysis does not tell us if tools akin to the conical top or the more simple bar used in ropemaking has been applied. Also, still to be answered is the question of whether the wires have been thermally treated
to enhance the moldability of the metal. Even so, the geometrical analysis reveals why the rings were repeatedly crafted with beautiful results.

\newpage
\begin{figure}[t]\centering
\caption{(A) Twisted neck rings (denoted R1-R5 counting from outside to inside) on display at The National Museum of Denmark. (B) Length $L(n)$ of twisted wire versus number of rotations plotted for two strands of diameter 5~mm and length 
50~cm. (C) $A_r$ versus pitch angle plotted for $N=2$ and 3, respectively. For pitch angles higher than that of the maximally rotated structure the curve is blue, below it is red. Circles indicate measurements on ten rings; six rings are clustered around $v_\bot =45^\circ$, two of which coincide (Table 2).}
\includegraphics[width=5.1cm]{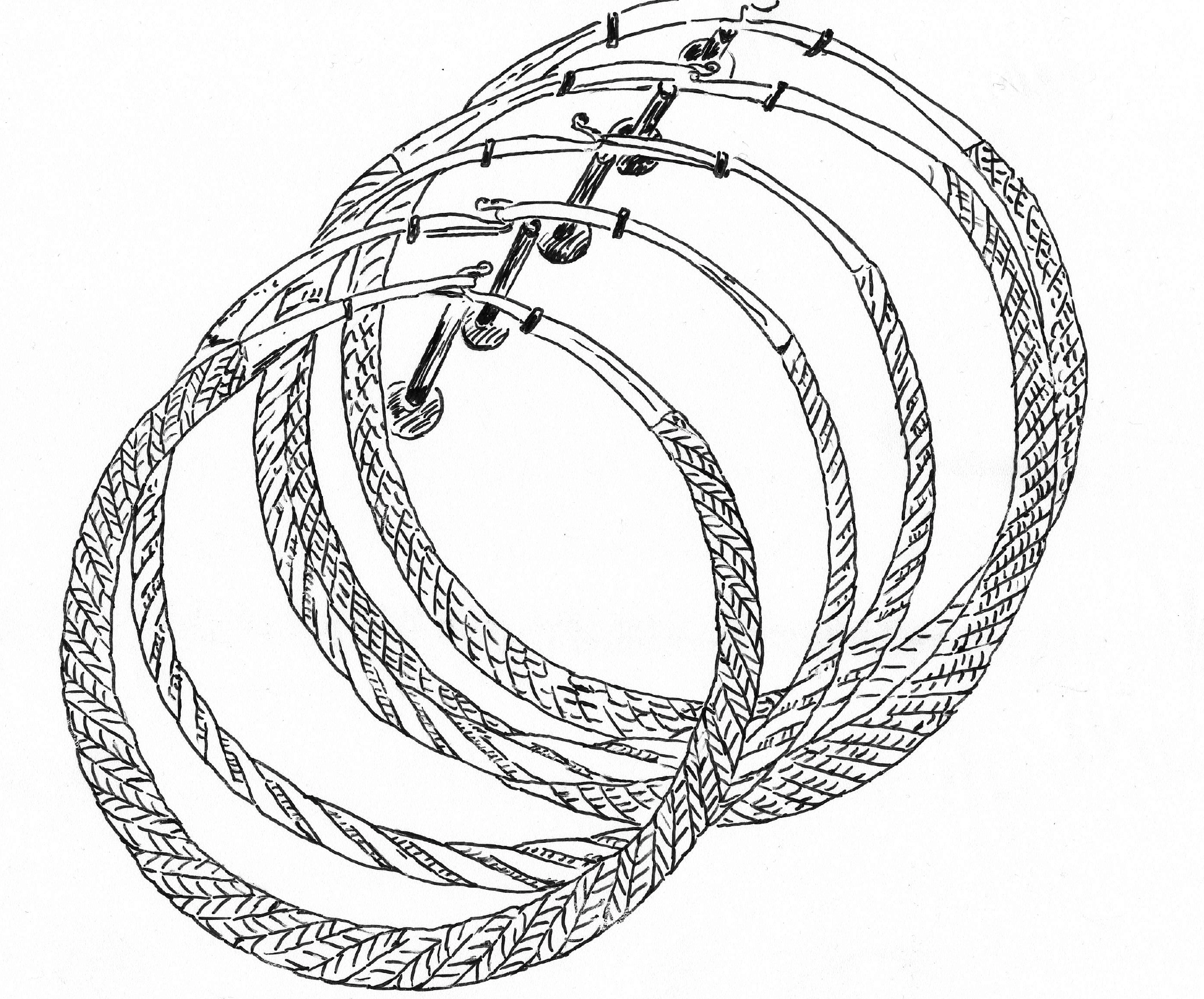}
\end{figure}
\begin{figure}[h]\centering
\includegraphics[width=5.1cm]{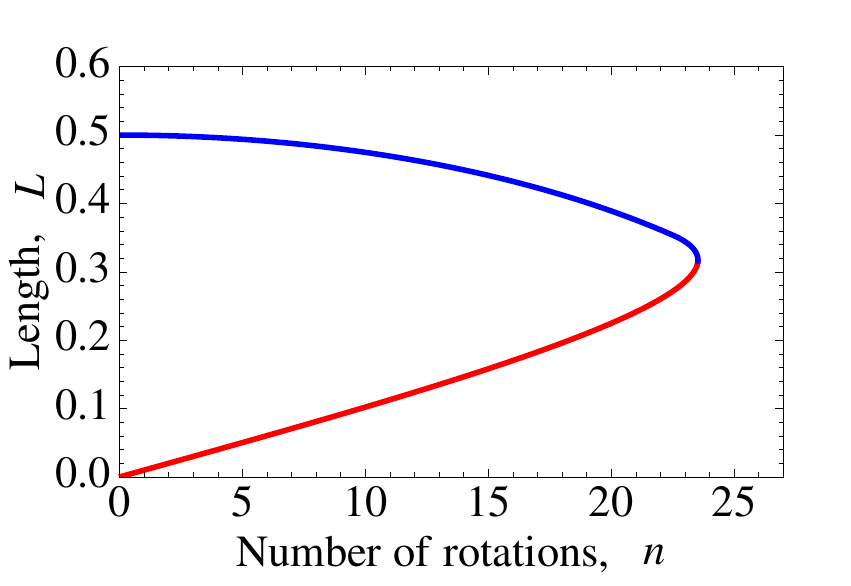}
\end{figure}
\begin{figure}[h]\centering
\includegraphics[width=5.1cm]{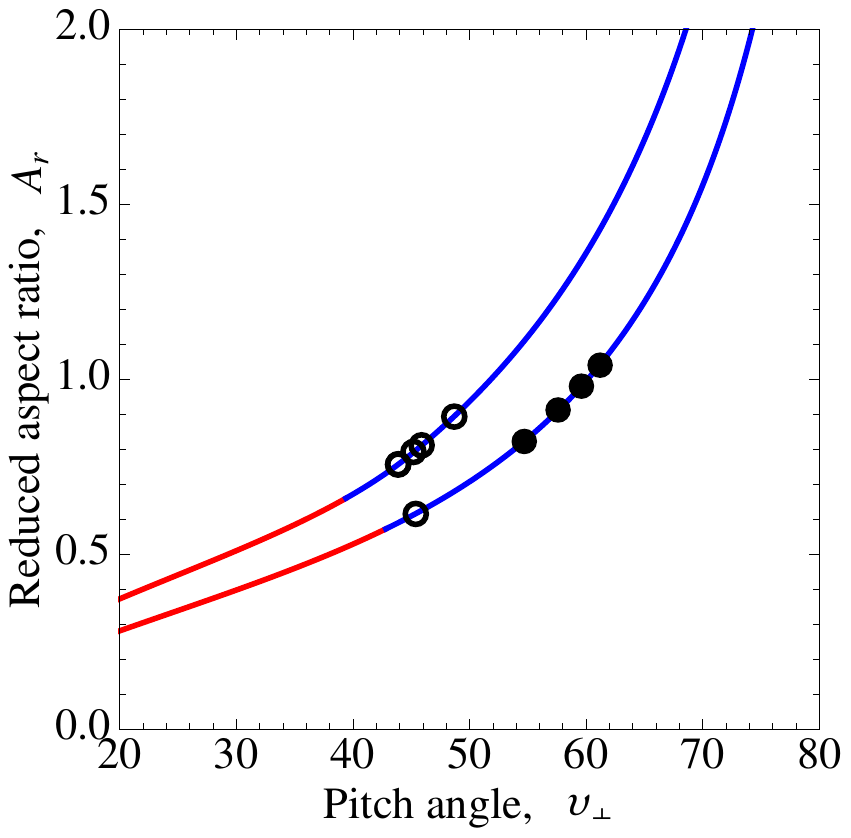}
\end{figure}

\newpage
\begin{table}[h]\centering
\label{tab:1}      
\begin{tabular}{clc}
\hline\noalign{\smallskip}
$N$ & $v_{\bot}$ & $A_r$\\
\noalign{\smallskip}\hline\noalign{\smallskip}
2 & 39.4$^\circ$ & 0.66\\
3 & 42.8$^\circ$ & 0.57\\
4 & 43.8$^\circ$ & 0.49\\
\noalign{\smallskip}\hline
\end{tabular}
\caption{The pitch angle, $v_\bot$, and the reduced aspect ratio $A_r=H/(ND)$ 
for the maximally rotated structures as a function of the number of strands, $N$.}
\end{table}\vspace{4cm}

\begin{table}[h]\centering
\label{tab:1}       \begin{tabular}{ccccc}
\hline\noalign{\smallskip}
Object (origin) & Pattern & $N$ & $A_r$  &$v_\bot$ \\
\noalign{\smallskip}\hline\noalign{\smallskip}
R1 (DNM) & $(i)$ & 2 & $^a$0.813 $^b$0.815 $^c$0.757  & 43.9$^\circ$ \\
R4 (DNM) &  $(i)$ & 3 & $^a$0.615 $^b$0.676  $^c$0.764 & 45.4$^\circ$ \\
R5  (DNM) & $(i)$ & 2 & $^a$0.887  $^b$0.756  $^c$0.886  & 43.9$^\circ$ \\
R6  (DNM, Tiss{\o}, Denmark) & $(i)$ & 2 & $^a$0.793 $^b$0.792 $^c$0.797  & 45.2$^\circ$ \\
R7 (DNM, Terslev, Denmark) & $(iii)$ & 3 & $^a$0.822  $^b$0.900 $^c$0.915 & 54.7$^\circ$ \\ 
R8  (DNM, Bonderup, Denmark) & $(i)$ & 2 &$^a$1.07 $^b$1.04 $^c$0.893 & 48.7$^\circ$ \\ \hline 
$\left[24,12\right]$  (JGC, Orkney) &  $(ii)$ & 3 & $^a$1.13 $^b$1.07 $^c$1.04 & 61.2$^\circ$ \\
$\left[24,15\right]$  (JGC, Orkney) & $(ii)$ & 3 & $^a$1.11 $^b$0.98 $^c$1.08  & 59.6$^\circ$ \\
$\left[24,21\right]$  (JGC, Orkney) & $(iii)$ & 3 & $^a$0.976 $^b$0.912 $^c$0.977  & 57.6$^\circ$ \\
$\left[24,22\right]$  (JGC, Orkney) & $(i)$ & 2 &  $^a$0.811 $^b$0.815 $^c$0.838  & 45.9$^\circ$ \\
\noalign{\smallskip}\hline
\end{tabular}
\caption{Measurements on viking arm and neck rings; R6 is of gold, the rest of silver. 
Each ring has a small variation in how tightly it is laid.
Therefore, the reduced aspect ratio $A_r=(H/ND)$ is measured at three different places on each ring (4th column). The pitch angle is calculated for the smallest value which corresponds to the most tightly laid part (5th column). DNM = National Museum of Denmark; JGC=James Graham-Campbell \cite{graham1995}. }
\end{table}

\end{document}